\documentclass{appolb}
\usepackage{graphicx}
\usepackage{ogonek}
\usepackage{amsmath}
\usepackage{bm}

\newcommand{\gras}[1]{\boldsymbol{#1}}

\begin{document}
\title{Strong-interaction isospin-symmetry breaking within the~density functional theory%
\thanks{Presented at the XXII Nuclear Physics Workshop, Kazimierz Dolny, Poland, September 22-27, 2015}%
}
\author{P.~B\k{a}czyk$^a$, J.~Dobaczewski$^{a,b,c,d}$, M.~Konieczka$^a$, W.~Satu\l{}a$^{a,b}$
\address{$^a$Faculty of Physics, University of Warsaw, Pasteura 5, PL-02-093 Warsaw, Poland\\
$^b$Helsinki Institute of Physics, P.O. Box 64, FI-00014 University of Helsinki, Finland\\
$^c$Department of Physics, P.O. Box 35 (YFL), FI-40014  University of Jyv\"askyl\"a, Finland\\
$^d$Department of Physics, University of York, Heslington, York YO10 5DD, UK}
}
\maketitle
\begin{abstract}
The conventional Skyrme interaction is generalized by adding zero-range charge-symmetry-breaking and charge-independence-breaking
terms, and the corresponding energy density functional is derived. It is shown that the extended model
accounts for experimental values of mirror and triplet displacement energies (MDEs and TDEs) in
$sd$-shell isospin triplets with, on average, $\sim$100~keV precision using only two additional adjustable coupling constants. Moreover, the model is
able to reproduce, for the first time, the $A=4n$ versus $A=4n+2$ staggering of the TDEs.
\end{abstract}

\section{Introduction}

Mean-field (MF) method based on isospin-invariant
Skyrme~\cite{[Sky59]} interaction is proven to be extremely
successful in reproducing bulk nuclear properties, see~\cite{[Ben03]}
and Refs.\ cited therein.  There is, however, a clear experimental evidence
that the strong nucleon-nucleon (\textit{NN}) interaction violates the isospin
symmetry. Based on the differences in phase shifts and scattering
lengths, it was shown that the \textit{nn} interaction is $\sim$1\%
stronger than \textit{pp} interaction and that the \textit{np}
interaction is $\sim$2.5\% stronger than the average of \textit{nn}
and \textit{pp} interactions~\cite{[Mac01a]}.

The Coulomb force plays very important role in the formation of
nuclear structure. At the same time, acting only between protons, it
is the main source of breaking of the isospin symmetry.  A systematic
study by Nolen and Schiffer~\cite{[Nol69]} showed that the experimental differences between the binding energies (BE) of the mirror nuclei, mirror displacement energies (MDEs):
\begin{equation}
\mathrm{MDE}=BE\left(T,T_z=-T\right)-BE\left(T,T_z=+T\right), \label{eq:MDE}
\end{equation}
cannot be reproduced with the Coulomb interaction as the only source of the isospin-symmetry breaking (ISB), see also~\cite{[Bro00b],[Kan13s],[Col98]}.
Another effect which cannot be reproduced by means of an approach involving only isoscalar strong force is the so-called triplet displacement energy (TDE)~\cite{[Sat14s]}:
\begin{eqnarray}
\mathrm{TDE}=BE\left(T=1,T_z=-1\right)+BE\left(T=1,T_z=+1\right)\notag\\
-2BE\left(T=1,T_z=0\right),\label{eq:TDE}
\end{eqnarray}
which measures the curvature of binding energies of isospin triplets. The MDEs and TDEs are related to the charge-symmetry breaking (CSB) and
charge-independence breaking (CIB) components of the \textit{NN} interaction, respectively. The aim of this work is to present the preliminary results of the generalized Skyrme approach that includes
the CSB and CIB zero-range terms and quantifies their impact on the MDEs and TDEs.

\section{Classification of the ISB interactions}

On a fundamental level, the isospin symmetry is broken due to (i)
different masses and electromagnetic interactions of $u$ and $d$
quarks (which translates at a hadronic level into differences of the
masses of hadrons within the same isospin multiplet), (ii) meson mixing,
and (iii) irreducible meson-photon exchanges. The CSB mostly originates
from the difference in masses of protons and neutrons, leading to
the difference in the kinetic energies and influencing the boson
exchange. For the CIB, the major cause is the pion mass splitting. For
more details see Refs.~\cite{[Mac01a],[Mil95]}.

Henley and Miller introduced a convenient and commonly used
classification of various ISB terms~\cite{[Mil95],[Hen79]}. According
to this classification, the isospin-invariant (isoscalar) \textit{NN}
interactions are called the class~I forces. The class~II isotensor
forces preserve the charge symmetry, breaking charge independence at the same time. The
class~III forces break both the charge independence and charge
symmetry, staying fully symmetric under interchange of nucleonic indices in
the isospace. Finally, forces of class~IV break both symmetries and
mix isospin already at the two-body level. The classification is commonly
used in the framework of models based on boson-exchange formalism,
like CD-Bonn~\cite{[Mac01a]} or AV18~\cite{[Wir13s]}. So far, apart from Ref.~\cite{[Bro00b]}, it has
not been directly used within the DFT formalism, which is usually based on 
isospin-invariant strong forces.

\section{Extended Skyrme model}

To account for the CIB and CSB effects, we have extended the conventional Skyrme interaction by adding zero-range interactions
of class II and class III:
\begin{eqnarray}
\hat{V}^{\rm{II}}(i,j) & = &
\frac12 t_0^{\rm{II}}\, \delta\left(\gras{r}_i - \gras{r}_j\right)
\left(1 - x_0^{\rm{II}}\,\hat P^\sigma_{ij}\right)
\left[3\hat{\tau}_3(i)\hat{\tau}_3(j)-\hat{\vec{\tau}}(i)\circ\hat{\vec{\tau}}(j)\right],
\label{eq:Skyrme_classII}\\
\hat{V}^{\rm{III}}(i,j) & = &
\frac12 t_0^{\rm{III}}\, \delta\left(\gras{r}_i - \gras{r}_j\right)
\left(1 - x_0^{\rm{III}}\,\hat P^\sigma_{ij}\right)
\left[\hat{\tau}_3(i)+\hat{\tau}_3(j)\right],
\label{eq:Skyrme_classIII}
\end{eqnarray}
where $t_0^{\rm{II}}$, $t_0^{\rm{III}}$, $x_0^{\rm{II}}$, and $x_0^{\rm{III}}$ are adjustable parameters and
$\hat P^\sigma_{ij}$  is the spin-exchange operator. The corresponding contributions to energy density functional (EDF)
read:
\begin{eqnarray}
\mathcal{H}_{\rm{II}} = & \frac{1}{2}t_0^{\rm{II}}\left(1-x_0^{\rm{II}}\right) & \hspace{-6pt}
(\rho_n^2+\rho_p^2-2\rho_n\rho_p-2\rho_{np}\rho_{pn}\notag\\
& &-\gras{s}_{n}^2-\gras{s}_{p}^2+2\gras{s}_{n}\cdot\gras{s}_{p}+2\gras{s}_{np}\cdot\gras{s}_{pn}),\label{eq:ED_classII}
\\
\mathcal{H}_{\rm{III}} = & \frac{1}{2}t_0^{\rm{III}}\left(1-x_0^{\rm{III}}\right) & \hspace{-6pt}
\left(\rho_n^2-\rho_p^2 - \gras{s}_{n}^2+\gras{s}_{p}^2\right),
\label{eq:ED_classIII}
\end{eqnarray}
where $\rho$ and $\gras{s}$ are scalar and spin (vector) densities, respectively. Note, that the effect of spin exchange
leads to a trivial rescaling of the coupling constants, and can be omitted by setting $x_0^{\rm{II}} = x_0^{\rm{III}}=0.$
Hence, the extended formalism depends on two new coupling constants.

The contribution to EDF from the class III force depends entirely on
the standard $nn$ and $pp$ densities and, therefore, can be taken into account
within the conventional $pn$-separable DFT approach. The contribution
from the class II force, on the other hand, depends explicitly on the
mixed densities, $\rho_{np}$ and $\gras{s}_{np}$, and requires the use
of $pn$-mixed DFT~\cite{[Sat13cs],[She14s]}, augmented by the isospin
projection to control this degree of freedom.

The proposed extension was implemented within the code
HFODD~\cite{[Sch12a]} that allows for the $pn$-mixing in the
particle-hole channel. The isospin degree of freedom is controlled
using the isocranking method -- an analogue of the cranking technique,
which is widely used in high-spin physics~\cite{[Sat13cs]}. The
method allows us to calculate the entire isospin multiplet, $T$, by
starting from an isospin-aligned state $|T,T_z=T\rangle$ and
isocranking it by an angle $\theta$ around the $x$-axis in the isospace.
The isocranking can be regarded as an approximate method to perform the isospin
projection. The rigorous treatment of the isospin quantum number
within the $pn$-mixed DFT formalism requires full, three-dimensional
isospin projection, which is currently under development.

\begin{figure}[!htb]
\begin{minipage}[b]{0.495\textwidth}
    \includegraphics[width=\textwidth]{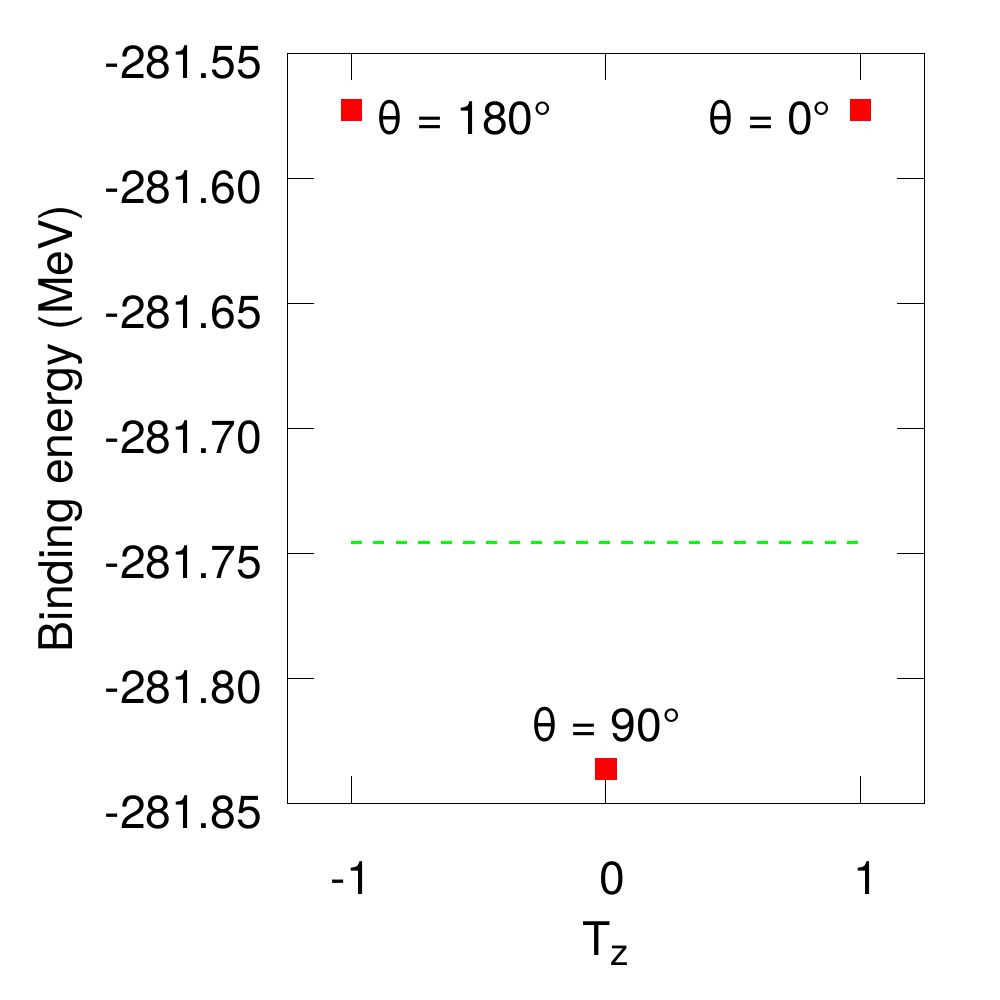}
\end{minipage}
\hfill
\begin{minipage}[b]{0.495\textwidth}
    \includegraphics[width=\textwidth]{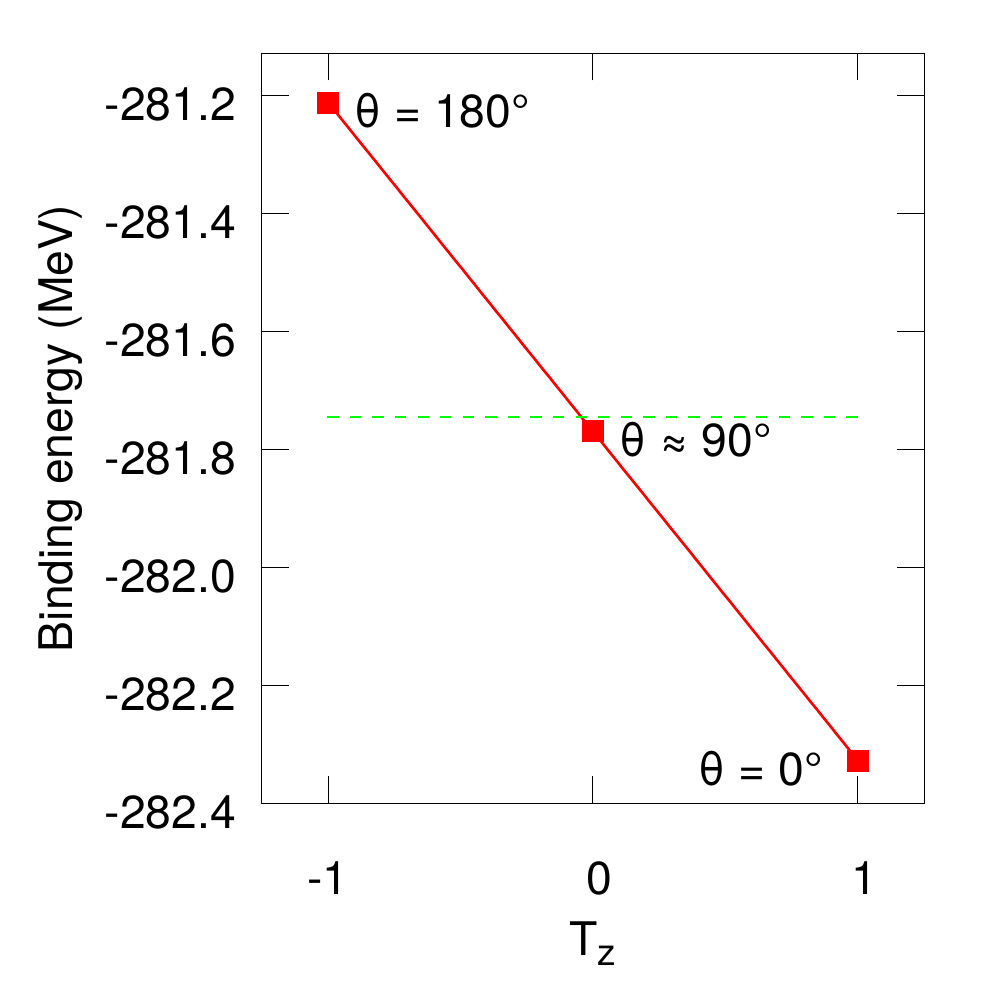}
\end{minipage}
\caption{Calculated g.s.\ energies of the $A=30$ isospin-triplet nuclei. Calculations were performed without Coulomb
interaction.
Full squares in the left and right panels show the results obtained using the class II and III forces, respectively.
The~dashed lines show the g.s.\
energies calculated without any ISB terms included. The solid line indicates an almost perfect linear trend of points
calculated with the class III force only.}
\label{fig:test}
\end{figure}

\section{Numerical results}

To investigate the influence of new terms on the
ground-state (g.s.) binding energies, we first performed a test calculation
without Coulomb for a~case of the isospin triplet in the
$A=30$ isobars. By adding to the isospin-invariant Skyrme
interaction either the class II or class III forces we were able to
delineate the influence of hadronic ISB forces on the binding
energies and TDE and MDE energy indicators. The results are depicted in
Fig.~\ref{fig:test}. As anticipated, the CIB class II force changes
the curvature (TDE) of binding energies  within the triplet but
almost does not affect the MDE of its $T_z=\pm1$ members. Conversely,
the class III force, which breaks the charge symmetry,
strongly affects the values of MDE, introducing only minor corrections to the
TDE, which are due to the self-consistency.

\noindent
\begin{figure}[!htb]
\begin{minipage}[b]{0.495\textwidth}
    \includegraphics[width=\textwidth]{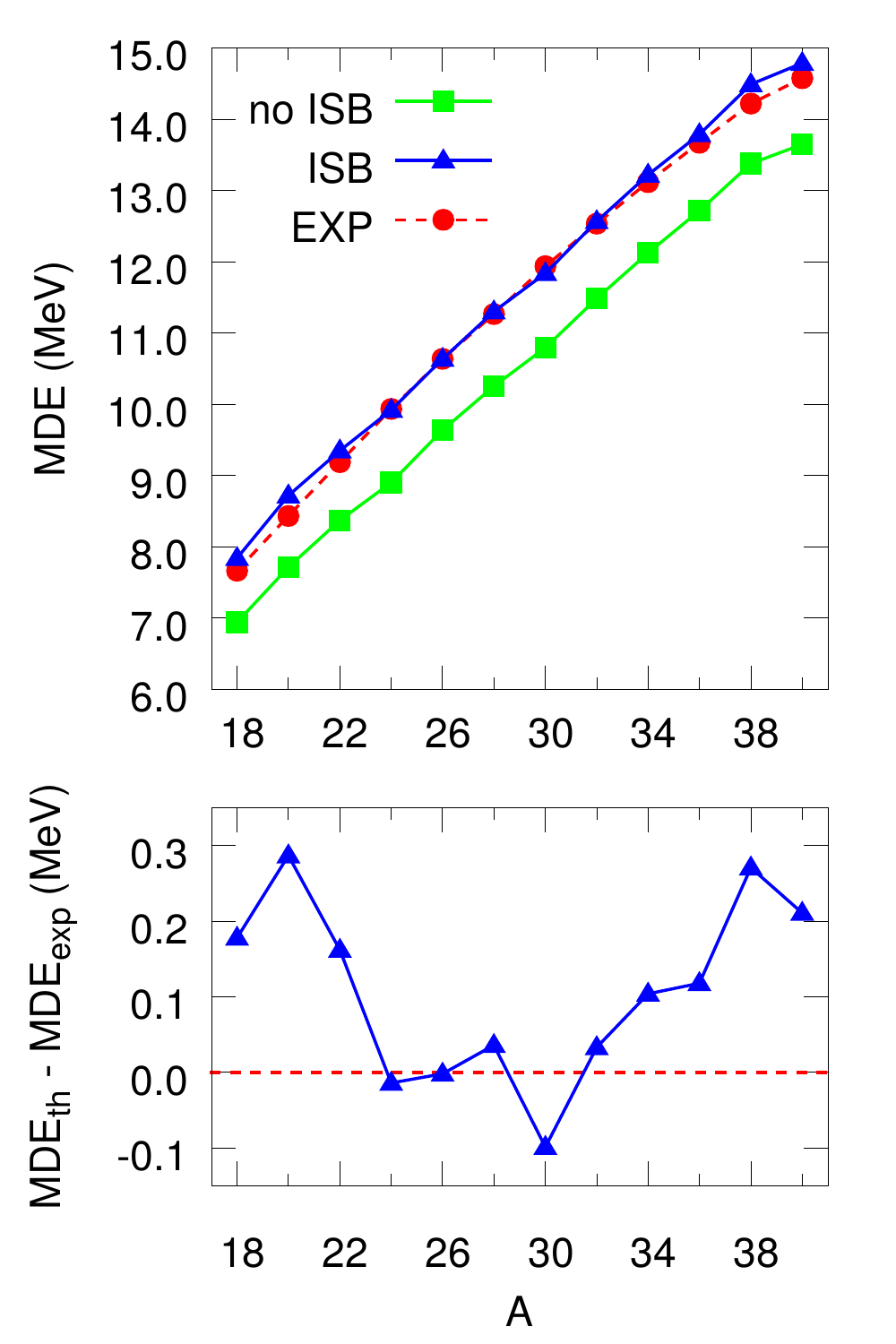}
\end{minipage}
\hfill
\begin{minipage}[b]{0.495\textwidth}
    \includegraphics[width=\textwidth]{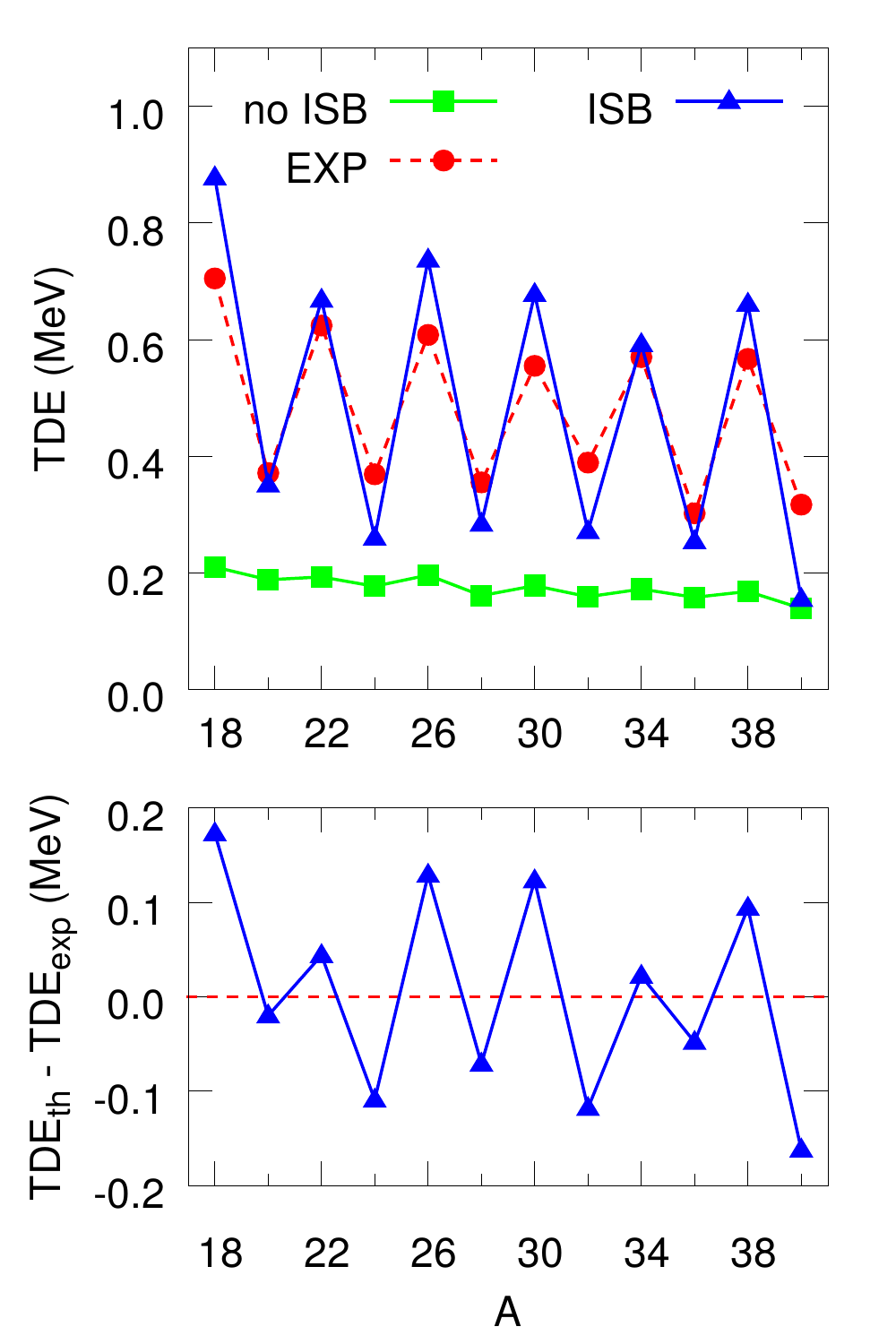}
\end{minipage}
\caption{The upper panels display the values of MDEs (left) and TDEs
(right) calculated for the isospin triplets in the $sd$-shell nuclei.
Circles show experimental points, squares represent results
of calculations involving isospin-invariant Skyrme force
SV~\cite{[Bei75s]} only, and triangles show results obtained
using the extended model with the hadronic ISB terms
(\ref{eq:Skyrme_classII}) and~(\ref{eq:Skyrme_classIII}) included.
Coulomb interaction was included. The lower panels show differences
between the theoretical calculations with the ISB terms included and
experimental values.
}
\label{fig:results}
\end{figure}

The test shows that the ISB forces of class II and III contribute
almost exclusively to TDEs and MDEs, respectively. It justifies our
strategy of fitting the $t_{0}^{\rm{II}}$ and $t_{0}^{\rm{III}}$
coupling constants to the TDE and MDE residuals $-$ the differences
between experimental and theoretical results obtained using
the conventional MF model that involves only the isospin-invariant Skyrme and Coulomb
forces. Moreover, since the residuals are relatively small, the fit
can be done in a perturbative way what leads to:
$t^{\rm{II}}_0=20$\,MeV and $t^{\rm{III}}_0=-8$\,MeV. These values
were subsequently used to calculate MDEs and TDEs for isospin
triplets in the $sd$-shell nuclei. The results are presented in
Fig.~\ref{fig:results}. Without the hadronic ISB forces, the
discrepancies between the experimental and the theoretical values of MDEs (dubbed the
Nolen-Schiffer anomaly~\cite{[Nol69]}) are of order of
1\,MeV. For TDEs, they are on average 0.3\,MeV. Moreover, the conventional model
cannot reproduce a very characteristic staggering of TDEs between the
$A=4n$ and $A=4n+2$ triplets.

The inclusion of the hadronic ISB terms of class II and class III
allows us to reduce the average disagreement between experiment and
theory to a level of about 100\,keV for TDEs and 130\,keV for MDEs. Moreover, as shown in the
figure, the extended model allows to account, for the first time, for
the $A=4n$ and $A=4n+2$ staggering of TDEs. It is worth underlying
that the results for the $4n+2$ triplets were obtained by
isocranking the isospin-aligned $|T=1,T_z=1\rangle$ MF solutions in
even-even nuclei, which are uniquely defined and represent the
$J=0^+$ ground states. The isospin-aligned $|T=1,T_z=1\rangle$ MF
solutions in the $4n$ triplets, on the other hand, refer to odd-odd
nuclei. These solutions are, in general, aligned in space and
represents the $J\ne 0$ states. Due to the shape-alignment ambiguity,
see Ref.~\cite{[Sat12s]}, the MF solutions in odd-odd nuclei are not
uniquely defined. The results shown in Fig.~\ref{fig:results}
represent arithmetic averages over the MF solutions that correspond
to spin alignments along the short, middle, and long axes of the
nuclear shape, respectively.

\section{Summary}

The conventional MF model involving the isospin-invariant Skyrme force
with Coulomb interaction included has been extended by adding two zero-range terms
that break charge symmetry and charge independence. The
two free parameters were adjusted to reproduce the experimental values of the MDEs
and TDEs.  This allowed us to reduce the discrepancy between experimental
and theoretical values to, on average, $\sim$100~keV, and to
reproduce, for the first time, the $A=4n$ and $A=4n+2$ staggering of the
TDEs. We plan to apply the extended model to study phenomena
sensitive to the isospin symmetry.

\vspace{0.3cm}

\noindent
This work was supported in part
by the Polish National Science Center under Contract Nos.\ 2012/07/B/ST2/03907 and 2014/15/N/ST2/03454,
by the ERANET-NuPNET grant SARFEN of the Polish National Centre for Research and Development (NCBiR), and
by the Academy of Finland and University of Jyv\"askyl\"a within the FIDIPRO program.
We acknowledge the CSC-IT Center for Science Ltd., Finland, for the allocation of
computational resources.

\bibliography{NOBEL,jacwit32}

\begin{thebibliography}{10}

\bibitem{[Sky59]}
{T.H.R. Skyrme, Nucl. Phys. {\bf 9}, 615 (1959)}.

\bibitem{[Ben03]}
{M. Bender, P.-H. Heenen, and P.-G. Reinhard, Rev. Mod. Phys. {\bf 75}, 121
  (2003)}.

\bibitem{[Mac01a]}
{R. Machleidt, Phys. Rev. C {\bf 63}, 024001 (2001)}.

\bibitem{[Nol69]}
{J.A. Nolen and J.P. Schiffer, Ann. Rev. Nuc. Sci. {\bf 19}, 471 (1969)}.

\bibitem{[Bro00b]}
{B.A. Brown, W.A. Richter, and R. Lindsay, Phys. Lett. {\bf B483}, 49 (2000)}.

\bibitem{[Kan13s]}
{K. Kaneko {\it et al.\/}, Phys. Rev. Lett. {\bf 110}, 172505 (2013).}

\bibitem{[Col98]}
{G. Col\`o {\it et al.\/}, Phys. Rev. C {\bf 57}, 3049 (1998)}.

\bibitem{[Sat14s]}
{W. Satu{\l}a {\it et al.\/}, Acta Phys. Polon. B{\bf 45}, 167 (2014).}

\bibitem{[Mil95]}
{G.A. Miller, and W.H.T. van Oers, \textit{Symmetries and Fundamental
  Interactions in Nuclei} edited by W.C. Haxton and E.M. Henley (World
  Scientific, Singapore, 1995).}

\bibitem{[Hen79]}
{E.M. Henley, and G.A. Miller, in \textit{Mesons in Nuclei}, edited by M. Rho
  and D.H. Wilkinson (North Holland, Amsterdam, 1979), p. 405.}

\bibitem{[Wir13s]}
{R.B. Wiringa {\it et al.}\/, Phys. Rev. C {\bf 88}, 044333 (2013).}

\bibitem{[Sat13cs]}
{K. Sato {\it et al.\/}, Phys. Rev. C {\bf 88}, 061301(R) (2013)}.

\bibitem{[She14s]}
{J.A. Sheikh {\it et al.\/}, Phys. Rev. C {\bf 89}, 054317 (2014)}.

\bibitem{[Sch12a]}
{N. Schunck {\it et al.}, Comput. Phys. Commun. {\bf 183}, 166 (2012)}.

\bibitem{[Bei75s]}
{M. Beiner {\it et al.\/}, Nucl. Phys. {\bf A238}, 29 (1975)}.

\bibitem{[Sat12s]}
{W. Satu{\l}a {\it et al.}, Phys. Rev. C {\bf 86}, 054316 (2012)}.

\end{thebibliography}
\bibliographystyle{unsrt}

\end{document}